\documentstyle[a4,german]{article}

\begin{document}

\title{Recovery of the Time-Evolution Equation of Time-Delay 
	Systems from Time Series}

\author{M. J. B\"unner, Th. Meyer, A. Kittel, and J. Parisi \\ 
	{\em Department of Energy and Semiconductor Research,} \\
	{\em Faculty of Physics, University of Oldenburg,} \\
	{\em D-26111 Oldenburg, Germany}
	\thanks{published in Phys. Rev. E {\bf 56} (1997) 5083}\\ 
	}

\date{February 13, 1997}

\large

\maketitle


\begin{abstract}

We present a method for time series analysis of both, scalar
and nonscalar time-delay 
systems. If the  dynamics of the system investigated is governed by 
a time-delay induced instability, the 
method allows to determine the delay time. In a second step, the
time-delay 
differential equation can be recovered from the 
time series.  The method is a generalization of 
our recently proposed method suitable
for time series analysis of {\it scalar} time-delay 
systems. The dynamics is not required to be settled on its 
attractor, which also makes transient motion accessible to the 
analysis. If the motion actually takes place on a 
chaotic attractor, the applicability of the method does 
not depend on the dimensionality of the chaotic attractor - one main
advantage 
over all time series analysis methods known until 
now. For demonstration, we analyze time series, 
which are obtained with the help of the numerical integration of a 
two-dimensional time-delay differential 
equation.   After having determined the delay 
time, we recover the nonscalar time-delay 
differential equation from the time series, in  agreement with the 
'original' time-delay equation.  Finally, 
possible applications of our analysis method in 
such different fields as medicine, hydrodynamics, laser 
physics, and chemistry are discussed.

P.A.C.S.: 05.45.+b
\end{abstract}


\section{Introduction}

Time-delay differential equations  have been widely proposed to 
account for the observed oscillatory, chaotic or hyperchaotic 
motion of dynamical systems. Most of the attention has been 
devoted to  scalar time-delayed models  \cite{mkg77} - 
\cite{tass95}. Since the pioneering work of Farmer \cite{farmer82}, 
it is well established that scalar time-delay differential equations are
able 
to exhibit high-dimensional chaotic attractors with many positive
Lyapunov 
exponents and, therefore, are prominent examples to illustrate the
chaotic 
hierarchy \cite{roessler83}. Since then, scalar time-delay equations,
especially the 
well-studied Mackey-Glass  system  \cite{mkg77}, have been used as model 
systems to produce high-dimensional chaotic time series. In 
the case of small delay times, the resulting low-dimensional chaotic
dynamics is accessible to time 
series analysis with the help of well-established methods 
\cite{grassberger83}-\cite{kantz94}. For example, the fractal dimension
and 
the Lyapunov exponents of the chaotic attractors  can be estimated. In
the case 
of large delay times, where the dynamics is high-dimensional 
chaotic, these methods run into severe problems \cite{hegger96}.

A first step towards the time series analysis of time-delay 
systems has been done by Fowler and Kember \cite{fowler93}, who  showed
how 
'smart embeddings' can indicate the 
presence of an underlying scalar time-delay system for any 
delay time. Later, we introduced a time series analysis method  in order
to 
verify the existence of an underlying time-delay system
\cite{buenner95}-\cite{ 
promotion}. If the dynamics is governed by a scalar time-delay 
differential equation, we have shown that the 
delay time and the time-delay differential equation can 
uniquely be  recovered from the time series. The method does not put any 
restriction on the dimensionality of the dynamics  analyzed, opening 
up a door towards the time series analysis of 
high-dimensional chaotic motion in time-delay systems.  Furthermore, the
method does 
not require the motion to be settled on its attractor. The 
method turned out to be practically insensitive against 
additional noise, hence providing a well-suited tool for 
the analysis of experimental time series. So far, we have  
successfully applied the method to time series taken from an electronic
oscillator 
\cite{buenner95} and  a computer experiment \cite{buenner96}.
Although some of the time-delay models investigated 
are indeed scalar ones, time-delay systems are,  in general,
nonscalar. Nonscalar time-delay differential equations  have 
been proposed in such different fields as arrays of coupled oscillators 
\cite{niebur91, nakamura94}, laser physics 
\cite{lang80}-\cite{fischer94}, physiology 
\cite{glass88} -\cite{foss96}, hydrodynamics 
\cite{villermaux95}, and chemistry \cite{khrustova95} to 
account for the observed unstable and chaotic dynamical behavior. 
Additionally, several models with multiple delay 
times have been investigated \cite{leberre86}-\cite{ikeda80}.
In this paper, we present a generalization of the time series 
analysis method proposed 
in  \cite{buenner95}-\cite{ buenner96}. Herewith, we are able to 
identify  nonscalar time-delay systems and, therefore, verify the
existence of an 
underlying time-delay induced instability. 
The method allows the recovery of the nonscalar 
time-delay differential equation from the time 
series. This article is organized as follows. In 
Section 2, the basic idea of 
the time series analysis method 
is presented. Also, adequate 
measures to determine the delay time from the time series are 
discussed. In Section 3, we illustrate the method by 
applying it to a two-dimensional time-delay system, the
trajectories of which are computed numerically.  
We show that, while the delay time can still be 
determined even with a scalar ansatz, 
the existence of an underlying time-delay system 
cannot be verified. The latter is accomplished with the help of a
nonscalar ansatz, 
which additionally allows to recover the time-delay equation from the
time series. 
Finally, possible applications to experimental systems are 
discussed in Section 4.

\section{The Basic Idea of the Analysis}

We consider  an $N$-dimensional time-delay 
differential equation
\begin{eqnarray}
\label{tdde}
\dot{\vec{y}}_0(t)  & = &  \vec{h}(\vec{y}_0(t), \vec{y}_{\tau_0}(t)),
\\
\vec{y}_{\tau_0}(t) & = & \vec{y}_0(t-\tau_0), \nonumber
\end{eqnarray}
with the initial condition
\begin{equation}
\vec{y}_0(t)=\vec{y}_{i}(t), \hspace{1.0cm} -\tau_0 \le t \le 0.  
\nonumber
\end{equation}
The time evolution of $\vec{y}_0$ at time $t$ does not 
only depend on its present state $\vec{y}_0(t)$, it 
also depends on a state in the past, $ \vec{y}_{\tau_0}(t)$, which
introduces nonlocal correlations in time. 
The state of the system is uniquely defined by $N$ 
functions on an interval of length 
$\tau_0$. Therefore, the phase space of system (\ref{tdde}) is 
${\cal{C}}_{\tau_0}^N$, where ${\cal C}_{\tau_0}$ is the space 
of continuous functions on the interval $[-\tau_0, 0]$ and the 
phase space has to be considered as being infinite dimensional.

The trajectory in the infinite dimensional phase space 
${\cal Y}(\tau,t) \in {\cal{C}}_{\tau_0}^N$ can be recovered
from its projection $\vec{y}_{0}(t)$ without loss of information
\begin{equation}
\label{trajectory}
{\cal Y}(\tau,t)=\vec{y}_{0}(t-\tau),\hspace{2.0cm} -\tau_0 \le \tau \le
0.
\end{equation}
We emphasize that the construction (\ref{trajectory}) of the trajectory
in 
phase space is exact and can be accomplished for all 
values of the control parameters of the time-delay equation. The 
$N$ time series  $\vec{y}_0(t)$ 
encompass the complete information about the dynamics of the 
system in the infinite dimensional phase space ${\cal 
C}_{\tau_0}^N$, no matter which dynamical 
state is realized by the time-delay system. Equation
(\ref{trajectory}) holds for transient motions as 
well as for motions on chaotic 
attractors of arbitrary dimension. In general, it 
is expected that the dynamics of an 
infinite dimensional system cannot be reconstructed from a 
finite number of time series. For instance, to construct the 
trajectory in the phase space of
spatial systems, the dynamics of which is governed by 
nonlinear partial differential equations, an increasing number of 
time series with  increasing 
complexity of the dynamics is required.

Additionally, the time derivative $\dot{{\cal Y}}(\tau,t)\in
{\cal{C}}_{\tau_0}^N$
of the trajectory in phase space can be estimated in principle
\begin{equation}
\label{zeitableitung}
\dot{{\cal Y}}(\tau,t)=\dot{\vec{y}}_{0}(t-\tau),\hspace{2.0cm} -\tau_0
\le \tau \le 0,
\end{equation}
though, in practice, the estimation of time derivatives
can be the source of severe errors.

In the case of  ordinary as well as time-delay 
differential equations, the time derivative of the
trajectory in phase space is functionally related to 
the trajectory in phase space via the time-evolution
equation. It is the specific property of time-delay
systems, though, that only a restricted number of coordinates
are correlated via the time-evolution equation (\ref{tdde}),
 namely, the $2N$ coordinates 
$(\vec{y}_0(t),\vec{y}_{\tau_0}(t))$ taken from the trajectory 
${\cal Y}(\tau,t)$ in phase
space with the $N$ coordinates $\dot{\vec{y}}_0(t)$ taken from the
time derivative $\dot{{\cal Y}}(\tau,t)$ of the trajectory.

The basic idea of the time series 
analysis method  presented is to test whether a time-delay 
equation (\ref{tdde}) can be constructed, the  
solution of which is given by the observed time 
series. In order to accomplish this task, in general, one has to 
construct a phase space with 
the help of the observed time series. In infinite 
dimensional systems, one would 
expect that an increasing  number of time series with 
increasing complexity of 
the dynamics  is required. We have 
argued in the preceding section that this is not 
true for time-delay systems, but 
the observation of  $N$ time series is 
sufficient to construct the 
trajectory in the infinite dimensional phase space of an 
$N$-dimensional time-delay system, no matter which dynamical 
state is realized. Secondly, it is the specific 
property of time-delay
systems that only a restricted number of coordinates
are correlated via the time-evolution equation (\ref{tdde}),
 namely, the $2N$ coordinates 
$(\vec{y}_0(t),\vec{y}_{\tau_0}(t))$ taken from the phase
space and one of its time derivatives $\dot{\vec{y}}_0(t)$.
Therefore, it is not necessary to analyze the dynamics
in the infinite dimensional phase space, to verify the
existence of an underlying time-delay system. It is 
sufficient to show the existence of a functional 
relationship (\ref{tdde}). To this end, we analyze the
dynamics in a $3N$-dimensional space, which is spanned
by the coordinates $(\vec{y}_0,\vec{y}_{\tau_0},\dot{\vec{y}_0})$.
The dynamics of a time-delay system in the $3N$-dimensional
space is restricted to a $2N$-dimensional hypersurface, 
which is given by the time-evolution equation (\ref{tdde}).
The hypersurface (\ref{tdde}) is defined that 
to any value  $(\vec{y}_0(t),\vec{y}_{\tau_0}(t))$ there is 
a unique value of the time derivative $\dot{\vec{y}_0}(t)$ 
for all times $t$.

The idea of the time series analysis method 
presented in this article is to test the existence 
of such a hypersurface for a given time series. 
Starting with $N$ scalar time series, which have been taken from 
the system to be investigated, we hypothesize, at first,  that the 
dynamics is governed by an $N$-dimensional 
time-delay system 
\begin{eqnarray}
\label{Nansatz}
\dot{\vec{y}}_0(t) & = &  \vec{h}_r(\vec{y}_0(t), \vec{y}_{\tau}(t)), \\
\vec{y}_{\tau}(t) & = &  \vec{y}_0(t-\tau), \nonumber
\end{eqnarray}
with an unknown function $\vec{h}_r$ and an unknown 
delay time $\tau$, both of which will be determined in
the subsequent analysis, if the ansatz (\ref{Nansatz})
turns out to be successful.
Then, we take the values of $(\vec{y}_0,\vec{y}_{\tau})$
and $\dot{\vec{y}_0}$ from the time series and analyze
its dynamics in the $3N$-dimensional space, which is
spanned by the coordinates 
$(\vec{y}_0,\vec{y}_{\tau},\dot{\vec{y}_0})$. 
If the coordinates of the trajectory
$(\vec{y}_0(t),\vec{y}_{\tau}(t),\dot{\vec{y}}_0(t))$ are 
functionally correlated via equation (\ref{Nansatz}), the hypothesis 
that the  system is governed by a time-delay 
equation with the delay time $\tau$ has been verified. If the projected
trajectory
$(\vec{y}_0,\vec{y}_{\tau},\dot{\vec{y}_0})$ 
does not fulfill condition (\ref{Nansatz}), the 
hypothesis has to be rejected. This is a unique criterion to 
determine the delay time from the time series. Additionally, the 
time-delay differential equation can be constructed by analyzing the 
functional relationship (\ref{Nansatz}), which exactly gives the 
function $\vec{h}_r$. Obviously, the only requirement remains 
that the system dynamics fulfills the time-evolution equation
(\ref{Nansatz}), 
which even is true for all kinds of transient motion as 
well as for the motion on chaotic or hyperchaotic attractors of
arbitrary dimension. 
Therefore, the method permits to 
analyze high-dimensional chaotic dynamics of 
time-retarded systems, which is not accessible to 
the fractal dimension analysis. 
If the system possesses several 
coexisting attractors \cite{losson93}, the method is applicable 
to the dynamics on every attractor. 

We have argued above that the existence of  an underlying 
time-delay system can be verified by proving 
the existence of a $2N$-dimensional hypersurface $\vec{h}_r$ in the
form (\ref{Nansatz}) in the $3N$-dimensional space, which is spanned
by the coordinates $(\vec{y}_0,\vec{y}_{\tau},\dot{\vec{y}_0})$.
Therefore, it is crucial to apply adequate 
measures, which enable us to identify such a hypersurface by 
analyzing  the time series. All measures proposed so far \cite{fowler93}
- 
\cite{buenner96} solely rely on the fact that, if the 
trajectory $(\vec{y}_0(t),\vec{y}_{\tau}(t),\dot{\vec{y}_0(t)})$ 
is correlated  via equation 
(\ref{Nansatz}), the dimensionality of the trajectory is reduced.   
To our knowledge, this has been realized the first time by 
Fowler and Kember \cite{fowler93}, who analyzed the dynamics of 
the Mackey-Glass equation. They applied an 
embedding of the time series in a three-dimensional space with 
two time-delayed coordinates. The delay time of the first coordinate has 
been chosen to be small. The delay time of the second coordinate was 
taken as variable. The authors of \cite{fowler93} stated that 
if the delay time of the second coordinate equals the delay time of 
the time-delay system, the 
trajectory lies 'close to a surface'. Fowler and 
Kember applied a singular value fraction to 
detect the decrease in dimensionality. As has been correctly 
mentioned by them, the 
singular value fraction is not a good tool, if the 
surface is folded. The latter  must be considered as 
the general case.

Recently \cite{buenner95}-\cite{promotion}, 
we have proposed a time series analysis method for scalar 
time-delay systems, only. There, we 
showed that the  trajectory in
the  $(y_0,y_{\tau_0},\dot{y}_0)$-space  is restricted to 
a two-dimensional surface. The reduction in dimension was detected by 
intersecting the projected trajectory with a 
surface $k(y_0,y_{\tau},\dot{y}_0)=0$. The intersection points 
must be on a curve, if the projected trajectory 
is correlated via  equation (\ref{Nansatz}). 
We have detected such a behavior by ordering the intersection points 
with respect to one coordinate and drawing a 
polygon line, which connects all intersection 
points. The length of the polygon line has been 
taken as a measure for the alignment of the 
intersection points. This simple method correctly 
determined the delay time of scalar 
time-delay systems, but there is no 
straightforward generalization for nonscalar 
time-delay systems. 

For this reason, we apply 
another method, the basic idea of which is 
the following: If the  trajectory of an $N$-dimensional
time-delay system in the 
$(\vec{y}_0,\vec{y}_{\tau_0},\dot{\vec{y}}_0)$-space,
where $\tau_0$ is assumed to be the correct value of the
delay time, is correlated  via equation (\ref{Nansatz}), 
the trajectory is restricted to a 
hypersurface. Therefore, most parts of the 
$(\vec{y}_0,\vec{y}_{\tau_0},\dot{\vec{y}}_0)$-space 
are not visited by the trajectory. 
If the trajectory is viewed in any other 
space, for instance, if the value of the delay time is not chosen
properly, 
the projected trajectory is expected to 
visit 'more' parts of the space.
Therefore, we compute the filling factor  of the 
projected trajectory by covering the 
$(\vec{y}_0,\vec{y}_{\tau},\dot{\vec{y}}_0)$-space with 
$P^{3N}$ equally sized hypercubes. The 
filling factor is the number of hypercubes, 
which are visited by the 
projected trajectory, normalized to the total 
number of hypercubes, $P^{3N}$. The filling factor is 
computed  under variation of $\tau$. The existence 
of an underlying time-delay induced instability induces 
a local minimum in the filling factor. 

Fowler and Kember \cite{fowler93} already 
suggested a fractal dimension analysis of the projected 
trajectories. The fractal dimension analysis has 
some severe drawbacks: At first, 
a fractal analysis requires a large number of data 
points, because for the determination of a 
fractal dimension it is crucial to resolve the geometrical 
object under investigation on different 'length scales'.
Secondly, the fractal analysis is computationally intensive and, 
in practice, sensitive to additional noise.  
All those measures rely on the fact that the 
trajectory is restricted to a 
hypersurface, if the projected trajectory is 
correlated  via equation (\ref{Nansatz}). It has 
been argued above that the dimension reduction is not a sufficient 
criterion for the verification of an underlying time-delay system. 
The existence of the functional relationship (\ref{Nansatz}) has to 
be shown separately.

\section{Time Series Analysis of a Two-Dimensional 
Time-Delay System - A Numerical Example}

The applicability of the method for time series analysis is 
demonstrated with the help of a computer experiment. A nonscalar 
time-delay system is integrated numerically. Details of the 
numerical integration are reported in the appendix. We will 
show that, although a scalar ansatz also leads to a local 
minimum in the filling factor, the scalar ansatz 
must be rejected, because it is not possible to find a 
surface, which is given by an equation of the form (\ref{Nansatz}) 
with the help of a 
scalar ansatz. In a second step, we will identify the system as 
a nonscalar time-delay system by 
verifying the existence of  such a surface with the help of 
a  nonscalar ansatz. Finally, the time-delay differential 
equation will be recovered from the time series.

We consider the two-dimensional time-delay 
differential equation, which has been chosen to serve its
demonstrational
purpose best:
\begin{eqnarray}
\label{uv}
\dot{u} & = & -  v + f(u_{\tau_0}), \\
\dot{v} & = &  g(u,v) ,  \nonumber
\end{eqnarray}
with the initial condition:
\begin{eqnarray}
u(t) & = & u_{i}(t),  \hspace{1.0cm} -\tau_0 
\le t \le 0, 
\nonumber \\
v(t=0) & = & v_{i}. 
\nonumber
\end{eqnarray}
The functions $f$ and $g$ are given by:
\begin{eqnarray}
f(u_{\tau_0}) & = & 
\frac{a u_{\tau_0}}{1+u_{\tau_0}^{10}}, \\
g(u,v) & = & - \frac{1}{T} (v - u).
\end{eqnarray}
Equation (\ref{uv}) has some similarity with the 
Mackey-Glass system. The dependence of 
$\dot{u}$ on the time-delayed value $u_{\tau_0}$  
is the same as it is the case in the Mackey-Glass 
system. But while the dependence of $\dot{u}$  on $u$ 
induces exponential relaxations in the Mackey-Glass system, in 
(\ref{uv}) it is similar to a damped oscillator.
There are two limits of the  control parameter space $(a, 
T,\tau_0)$, where  the dynamics of (\ref{uv}) is 
well-known. For $a \rightarrow 0$, system (\ref{uv}) 
reduces to a damped harmonic oscillator. For   $T  \rightarrow 0$, the
time 
scale of $v$ is much faster compared to the 
time scale of $u$. The variable $v$, then,  
adiabatically follows variable $u$ and the 
dynamics of   (\ref{uv}) resembles that 
of the Mackey-Glass system. Here, 
$a$ and $\tau_0$ are chosen such that in the scalar limit ($T  
\rightarrow 0$) the dynamics is 
high-dimensional chaotic ($a=3, \tau_0=20$). $T$ 
is varied from $0.10$ to $1.90$. Note that the system
\ref{uv} can be transformed to a scalar integro-differential
equation for the variable $u$ by integrating the second
equation with the help of the method of varying coefficients.

We present 
three time series of $u$ and $v$ for different $T$ in Fig. 1. In Fig. 
1(a), it is clearly seen how the variable $v$, 
for $T=0.10$, follows the variable $u$ and the dynamics of $u$ 
resembles that of a Mackey-Glass system. 
The values of $(u,v)$ are positive for all times. We mention 
at this point that system (\ref{uv})
is invariant under the transformation $(u,v) 
\rightarrow (-u, -v)$. Therefore, there exists 
another  attractor with negative values of  $(u,v)$. In Fig. 1(b) and 
Fig. 
1(c), we observe that the two coexisting attractors are 
merged. Variable $v$ no longer follows variable $u$, but it 
develops an independent dynamics. In these 
cases, the system reveals its nonscalar nature. 

Now, we analyze these time series
with the help of a filling factor analysis. At first, we choose 
the scalar ansatz  
\begin{equation}
\label{scalaransatz}
\dot{u}= h_r(u,u_{\tau}),
\end{equation}
for the analysis of the time series $u(t)$ with an  unknown 
delay time $\tau$ and an  unknown function $h_r$.We analyze the time
series in a 
three-dimensional space, which is spanned by the coordinates 
$(u, u_{\tau},\dot{u})$ with a variable value of $\tau$. Then, the
filling 
factor of the time series is determined 
under variation of $\tau$. 

The results of the filling 
factor analysis are presented in  Fig. 2 for different 
values of $T$.  The filling factor is minimal for 
small values of $\tau$ as a result of local correlations in 
time. The filling factor increases for 
increasing $\tau$, which is a fingerprint of the 
chaotic nature of the motion, and eventually 
reaches a maximal value. For $\tau=\tau_0=20.00$, a local 
minimum of the filling factor is observed for all 
values of $T$. This decrease in the filling factor 
is due to the nonlocal correlations in time 
induced by the time delay. An additional local 
minimum appears at $\tau=2\tau_0$.  For high 
enough values of $T$, though, other regularly spaced local 
minima in the filling factor appear, the period of which
is equal to the oscillations of the underlying damped oscillator. 
In Fig. 2(b), a blow-up of the $\tau$-dependent filling factor in the
vicinity of 
the delay time $\tau_0=20.00$ is shown. 
Clearly, the local minimum appears for all values of $T$ 
considered here, but it is less pronounced for 
increasing $T$, because the character of the time-delay 
system is more and more becoming nonscalar, then.  As has been
emphasized 
above, the reduction in dimension is only a 
necessary, but not a sufficient condition for the verification 
of a scalar time-delay system. The 
existence of a surface in the form (\ref{scalaransatz}) has to be 
checked as well. 

To this end, we apply an 
intersection of the time series $u(t)$ with the plane  
$\dot{u}(t^i)=0$. The values of the coordinates $u^i=u(t^i)$ and 
$u_{\tau_0}^i=u_{\tau_0}(t^ i)$ are recorded. They have to be 
correlated according to the scalar ansatz (\ref{scalaransatz}): 
\begin{equation}
h_r (u^i,u_{\tau_o}^i)=0.
\end{equation} 
Therefore, the points $(u^i,u_{\tau_o}^i)$ have to lie on a 
curve, if the dynamics is governed by 
a scalar time-delay equation. The intersection points $ 
(u^i,u_{\tau_o}^i)$ for $\tau=\tau_0=20.00$  and different values of 
$T$ are shown in Fig. 3.   For small values of $T$, the time 
series $v(t)$ follows the time series $u(t)$. In this case, 
the dynamics of variable $u$ is close to the dynamics  of the  
Mackey-Glass system. Such a behavior can be seen in Fig. 
3(a), where the intersection yields a geometrical 
object, which is close to being a one-dimensional 
curve. The inset of Fig. 3(a) shows a blow-up of the 
intersection points. Clearly, the alignment 
of the intersection points is not perfect, as a result of the 
nonscalar nature of system 
(\ref{uv}). Nevertheless, the scalar ansatz 
(\ref{scalaransatz}) would be a good approximation 
for small values of $T$.  For higher values of $T$, system 
(\ref{uv}) reveals its nonscalar nature. Therefore, 
in Fig. 3(b) and Fig. 3(c), the distribution of
the intersection points becomes cloudy. That means, a smooth functional 
relationship  
(\ref{scalaransatz}) cannot be found in these cases and the scalar 
ansatz has to be rejected. Nevertheless, in the considered example of
a two-dimensional time-delay system the
filling factor analysis was successful in the framework of a scalar
ansatz,
in the sense that the $\tau$-dependent filling factor showed local
minima
for the correct values of the delay time.

Now, we will analyze the time 
series for higher values of $T$ with the help of a 
nonscalar ansatz. The general two-dimensional ansatz is
\begin{eqnarray}
\label{nonscalaransatz2}
\dot{u} & = & h_{r,1}(u,u_{\tau},v,v_{\tau}), \\
\dot{v} & = & h_{r,2}(u,u_{\tau},v,v_{\tau}),
\end{eqnarray}
with an unknown delay time $\tau$ and two unknown functions
$h_{r,1},h_{r,2}$.
The analysis has to be conducted in a six-dimensional space, in which
the
dynamics is restricted to a four-dimensional hypersurface. The
hypersurface
is given by the functions  $h_{r,1},h_{r,2}$, which 
can be determined with the help of adequate fitting procedures, for
instance,
a least-squares-fit in the framework of a presupposed model.
In this article, we chose a more restrictive ansatz for demonstrational
purposes,
\begin{eqnarray}
\label{nonscalaransatz3}
\dot{u} & = & - v +f_r(u_{\tau}), \\
\label{nonscalaransatz4}
\dot{v} & = &   - g_r(u,v).
\end{eqnarray}
The delay time $\tau$ and the functions $f_r$ and $g_r$ are 
yet unknown and will be determined in the following.
At first, we perform a filling factor analysis in 
the same spirit as has been done in the scalar 
case. But now the two time series $(u(t),v(t))$ 
have to be projected to a six-dimensional space 
which is spanned by the coordinates $(u, 
u_{\tau},\dot{u},v, v_{\tau},\dot{v})$. The 
six-dimensional space is covered with equally sized hypercubes 
and the number of hypercubes which have been visited by the  
trajectory is counted under variation of $\tau$. 

The results are presented in Fig. 4 for different values of $T$, 
namely $T= 0.10, 0.60, 1.90$. The minimum in 
the filling factor for $\tau=\tau_0=20.00$ is well 
detected for all values of $T$. In Fig. 4(b), we 
show a blow-up of the $\tau$-dependent filling 
factor in the vicinity of the delay time $\tau_0$ 
of  system (\ref{uv}). Clearly, the local 
minimum for $\tau=\tau_0$ is detected for all values 
of $T$.   

As has been argued above,  the existence of the 
functional relationship 
(\ref{Nansatz}) has to be shown, in order to verify the underlying 
time-delay induced instability. The special form of 
ansatz (\ref{nonscalaransatz3})-(\ref{nonscalaransatz4}) together with
the 
time-evolution equations (\ref{uv}) allows for a 
convenient way of proving the existence of  the function
(\ref{Nansatz}). 
We emphasize, though, that, in general, it is expected to be more 
troublesome. We apply an intersection with the help 
of the condition $v(t^i)=0$. If the nonscalar ansatz 
(\ref{nonscalaransatz3})-(\ref{nonscalaransatz4}) is successful,  the
values 
$\dot{u}^i = \dot{u}(t^i)$ and $ u_{\tau_0}^i=u_{\tau_0}(t^i)$ 
have to be  correlated via
\begin{equation}
\label{fr}
\dot{u}^i =  f_r(u_{\tau_0}^i).
\end{equation}
Plotting $\dot{u}^i$ versus $u_{\tau_0}^i$ as is shown 
in Fig. 5(a), the existence of the smooth function $f_r$ 
is verified. We compare the reconstructed 
function $f_r$ (open circles) with the function $f$ 
(line) of the time-delay differential equation (\ref{uv}). The 
coincidence is good and the $(\dot{u}^i,u_{\tau_0}^i)$-plot can be used
to 
recover the function $f$ from the time series. We 
emphasize that no parameter has been adjusted to compare the function
$f$ with its recovery $f_r$ in Fig. 5(a).  

In  the next step, the functional relationship between 
$\dot{v}^i$ and $u^i$ is investigated. We use the same
intersection condition $v(t^i)=0$ as above. According to 
the nonscalar ansatz (\ref{nonscalaransatz3})-(\ref{nonscalaransatz4}),
the
coordinates of the intersection points are 
correlated via  
\begin{equation}
\label{gr}
\dot{v}^i  =    - g_r(u^i,0).
\end{equation}
Plotting $\dot{v}^i$ versus $u^i$  yields the function 
$g_r(u^i,0)$. We compare the functions $g(u^i,0)$ and $g_r(u^i,0)$ in
Fig. 5(b). Again, the 
correspondence is good and the  
$(\dot{v}^i,u^i)$-plot can be used to recover the function  
 $g_r(u^i,0)$ from the time series. 
The recovery of the function  $g_r(0,v^i)$ is to 
be done in the same spirit 
and is not shown here. Obviously, the existence of the 
functions $f_r$ and $g_r$  has been proven and 
a two-dimensional time-delay system has been identified
by analyzing the time series. 

\section{Applications}

Finally, we would like to discuss possible applications of the 
present method for time series 
analysis. We emphasize that our method does not have the 
restrictions which are inevitable to the 
embedding techniques necessary for the determination of fractal
dimensions 
of chaotic attractors in phase space. The analysis is not restricted to 
a low-dimensional chaotic motion. Transients can be 
analyzed as well. The method is not sensitive to additional 
noise. Furthermore, it has been shown that
the analysis can be performed with a comparably 
small number of data 
points \cite{promotion}. Apparently, we have a well-suited tool for 
the analysis of experimental time series. If 
the dynamics of the system to be investigated  is governed by a 
time-delay induced instability, the 
method allows for the determination of the delay time and a 
recovery of the time-delay differential 
equation. Therefore, it is possible to compare 
time series of experimental systems with 
proposed model equations in detail. System parameters can be 
extracted from the time series analysis, which 
might be not accessible otherwise. 

We speculate 
that the analysis can be particularly  useful in such fields as
medicine and biology, where noninvasive techniques 
are of great importance for obvious reasons. 
In several experiments on human subjects, which 
were exposed to time delays of some sort, a qualitative change 
of the observed dynamics has been verified 
\cite{mkg77, beuter93, tass95, glass88, beuter89}, which could 
be correlated to well-established pathologies. If the observed 
dynamics is, indeed, induced by a time delay, as proposed, 
we expect our analysis method to be successful. On the other hand, 
we propose to check the validity
of the Mackey-Glass system by analyzing suitable time series.The 
analysis might improve the 
understanding of the experiments and possibly 
allow to determine important system parameters and 
serve as a new diagnostic tool.  

In a recent article,  Villermaux 
\cite{villermaux95} deals with the low-frequency oscillations 
of the velocity field in the 
'hard turbulence' regime in a closed convection box. The author proposes
a 
two-dimensional time-delay system to describe the dynamics of the
disturbances. 
Our method has the potential to test the validity of the model 
by analyzing the experimental time series, which possibly can lead
 to a better understanding of boundary instabilities.

Another important class of a prototype model for chaotic
dynamics are laser systems.
 No wonder that time-delay models have also been investigated in laser
physics 
(\cite{ikeda87,lang80, arecchi91, leberre86, ikeda80} and references
therein).  It is the 
advantage of laser systems with a time delay that 
they allow precise measurements of time series \cite{ 
fischer94, gibbs81}. We find them
particularly suitable for our analysis, because we expect the 
time delay of laser systems to be practically discrete, 
compared to other experimental systems. The first steps towards the
identification of high-dimensional chaotic dynamics of a time-delayed
laser
system has been taken \cite{promotion}.

\section{Concluding Remarks}

In conclusion, we have presented a generalization of a recently 
proposed method for recovering 
the time-evolution equation of scalar time-delay systems by analyzing
the 
time series. The method is generalized in the way that it can be applied 
to nonscalar time-delay systems. 

We have shown that an $N$-dimensional time-delay system
can be identified with the help of $N$ time series. The analysis
has not to be conducted in the infinite dimensional phase space,
instead it is sufficient to analyze the dynamics in a $3N$-dimensional
space, in which the dynamics has to be restricted to a $2N$-dimensional
hypersurface. This finding gives us a unique 
criterion to determine the delay time of a 
nonscalar time-delay system by analyzing the time 
series. Additionally, the time-delay 
differential equation of dimension $N$ can be recovered. 

We emphasize that we only require the 
motion to obey the time-delay differential equation. The motion 
is not required to be located in certain 
parts of phase space. If the dynamical system 
possesses coexisting attractors, the method can 
be applied to motions on every coexisting attractor. 
Moreover, the method is also applicable 
to transient motions. If the motion is on a 
chaotic or hyperchaotic attractor, the applicability 
of the analysis method does neither depend on the 
dimensionality nor on the number of positive 
Lyapunov exponents of the  chaotic or hyperchaotic attractor. 
Therefore, we find that the present method might open up a 
door towards the time series analysis of high-dimensional chaotic motion
in 
time-delay systems.  

We have shown the applicability of the method by 
analyzing  time series which have been obtained 
with the help of numerically integrating  
a two-dimensional time-delay differential 
equation. The system  investigated mimics a scalar 
time-delay system in a certain parameter range, 
where a scalar ansatz yields a good 
approximation. Under variation of a single control 
parameter, the dynamics increasingly reveals 
the nonscalar nature of the time-evolution 
equation. In this case, a scalar ansatz is not 
sufficient, but, nevertheless, we have 
successfully analyzed the time series with the help of a 
nonscalar ansatz. The delay time has been determined from the 
time series. Finally, we recovered 
the two-dimensional time-delay differential 
equation by analyzing  the time series. Possible 
applications for the analysis of dynamical 
systems in such different fields as medicine, 
hydrodynamics, laser physics, and chemistry are 
discussed.

\section{Acknowledgements}

We thankfully acknowledge valuable discussions 
with J. Peinke, O. E. R\"ossler, the ENGADYN group and H. Kantz. 
Financial support 
of the Deutsche Forschungsgemeinschaft is acknowledged.

\newpage

\section{Appendix: Numerical Methods}

The results presented in this article have been computed with 
the help of a Runge-Kutta algorithm of fourth order.
The fundamental time step was taken to be $0.01$. The length of 
the memory  has been $2,000$ time steps. In all 
simulations presented, we have 
chosen homogeneous intial values $u(t)=0.50, v(t)=0.90, -\tau_0 \le t
\le 
0$. 

We checked the validity  of 
the numerical integration of  system (\ref{uv}) 
by comparing  the results of 
Runge-Kutta algorithms of different order, under 
variation of the time step of the integration, in order to check 
the validity of the results. Additionally, the computed 
time series were compared to analytical solutions, which are 
available in certain parameter ranges.

For the filling factor analysis, we used time 
series with $1,000,000$ data points. The time 
derivatives were estimated by applying a local 
parabolic approximation. To compute the $\tau$-dependent filling factor, 
every tenth point of the time series was taken. Therefore, we conducted 
the filling factor analysis by analyzing $100,000$ data points in the
three-dimensional or six-dimensional space, respectively. 

\newpage

\section*{Figure captions}
\begin{itemize}

\item[{\bf Fig. 1:}]
Time series of the system (\ref{uv}) obtained with the help of 
numerical integration ($ a=3.00, \tau_0=20.00$) for different 
values of $T$: (a) $T=0.10$; (b) $T=0.60$; (c) $T=1.90$. 

\item[{\bf Fig. 2:}] 
(a) Filling factor of the time series $u(t)$ of the system 
(\ref{uv}) with  $ a=3.00, \tau_0=20.00$ 
for different values of $T$ {\it (lower curve:} $T=0.10;$  
{\it middle curve:} $0.60;$ {\it upper curve:} $1.90$). We used 
$100,000$ data 
points for the filling factor analysis, which were taken out of a time 
series of $1,000,000$ data points. In (b) a blow-up 
of the filling factor in the vicinity of the delay 
time $\tau =\tau_0$ is shown $(squares: T=0.10; 
circles: T=0.60; stars: T=1.90)$.

\item[{\bf Fig. 3:}] 
Intersection points of the time series $u(t)$ for $ a=3.00, 
\tau_0=20.00$ and different values of 
$T$: (a) $T=0.10$, (b) $T=0.60$,  (c) $T= 1.90$. The 
analysis has been conducted with a time series 
of  $1,000,000$ data points.

\item[{\bf Fig. 4:}]
(a) Filling factor of the two time series $(u(t),v(t))$ of the 
system (\ref{uv}) with  $ a=3.00, \tau_0=20.00$ for different values of
$T (T=0.10, 
0.60, 1.90$; the values of $T$ are indicated in the figure). We used  
$100,000$ data points for 
the filling factor analysis, which were taken out 
of a time series of $1,000,000$ data points.  In 
(b) a blow-up of the filling factor in the vicinity of 
the delay time $\tau =\tau_0$ is shown $(squares: 
T=0.10; circles: T=0.60; stars: T=1.90)$.

\item[{\bf Fig. 5:}]
Recovery of the time-delay differential equation from the time 
series: (a) Comparison of the data 
points $(\dot{u}^i, u_{\tau_0}^ i)$, which are 
shown as open circles, with the function $f$ 
(line). For clarity, only $300$ data points 
$(\dot{u}^i, u_{\tau_0}^ i)$ are shown. (b) Comparison 
of the data points $(\dot{v}^i, u^ i)$, which are 
shown as open circles, with the function 
$g(u^i,0)$ (line). For clarity, only $40$ data points 
$(\dot{v}^i, u^ i)$ are shown.

\end{itemize}

\end{document}